\documentclass[journal,twoside,web]{ieeecolor}

\usepackage{generic}
\usepackage{cite}
\usepackage{amsmath,amssymb,amsfonts}
\usepackage{graphicx}
\usepackage{textcomp}

\usepackage{epsfig} 
\usepackage{lcsys}

\usepackage{xurl}
\usepackage[colorlinks=true, urlcolor=blue]{hyperref}
\usepackage{todonotes}
\usepackage{tikz}
\usepackage{pgfplots}
\usepgfplotslibrary{statistics,groupplots}

\usepackage{graphics,graphicx,psfrag,array,eepic}
\usepackage{multirow}
\usepackage{hhline}
\usepackage{mathtools}
\usepackage[strict]{changepage}
\usepackage{epsfig}
\usepackage{amsmath}
\usepackage{amssymb}
\usepackage{float}
\restylefloat{table}
\usepackage{multicol}
\usepackage{color}
\usepackage{balance}
\usepackage{cite}
\usepackage{booktabs} 
\usepackage{caption,booktabs}
\usepackage[tableposition=top]{caption}
\usepackage{subcaption}
\usepackage{lipsum}
\usepackage{comment}
\usepackage{amssymb}
\usepackage{graphicx}
\usepackage{comment}
\usepackage{bm}
\usepackage{xfrac}
\usepackage{algorithm}
\usepackage[noend]{algpseudocode}
\usepackage{mathtools, cuted}
\usepackage[thinc]{esdiff}
\usepackage{psfrag}
\usepackage{array}
\usepackage{latexsym,theorem}
\usepackage{amsfonts,amssymb,amsbsy}
\usepackage{tikz}
\usetikzlibrary{decorations.shapes,shapes.geometric,shadows,arrows,automata,positioning,calendar,mindmap,backgrounds,scopes,chains,er,3d,matrix,fit}
\usepackage{tikz-cd}
\usepackage{pgfplots}
\usetikzlibrary{intersections, pgfplots.fillbetween}
\DeclareMathOperator*{\argmin}{argmin}

\newcommand{\R}{{\rm I\!R}}

\newcommand{\N}{{\rm I\!N}}

\newcommand{\cO}{\mathcal{O}}

\newcommand{\cU}{\mathcal{U}}

\newcommand{\cE}{\mathcal{E}}

\newcommand{\cT}{\mathcal{T}}
\newcommand{\cK}{\mathcal{K}}
\newcommand{\cM}{\mathcal{M}}

\newcommand{\cL}{\mathcal{L}}
\newcommand{\cV}{\mathcal{V}}

\newcommand{\cH}{\mathcal{H}}

\newcommand{\cI}{\mathcal{I}}

\newcommand{\cG}{\mathcal{G}}

\newtheorem{definition}{Definition}
\newtheorem{theorem}{Theorem}
\newtheorem{proposition}{Proposition}

\newtheorem{assumption}{Assumption}

\usepackage[british]{babel}
\usepackage{mathrsfs}
\usepackage{url}
\usepackage{bm}
\usepackage{nicefrac}
\allowdisplaybreaks
\hypersetup{colorlinks=true, urlcolor=blue}
\usepackage{color}

\DeclareRobustCommand{\pentagon}{%
\tikz[baseline=-0.6ex,scale=0.15]{
\draw (90:1) -- (18:1) -- (-54:1) -- (-126:1) -- (162:1) -- cycle;
}}
\usepackage{alphalph,etoolbox}
\usepackage{mathtools,subcaption,float}
 \usepackage{caption}
\allowdisplaybreaks
\usepackage{pgfplots}
\usepgfplotslibrary{colorbrewer}
\pgfplotsset{compat = 1.15} 
\usetikzlibrary{pgfplots.statistics, pgfplots.colorbrewer} 

\begin{document}

\def\BibTeX{{\rm B\kern-.05em{\sc i\kern-.025em b}\kern-.08em
    T\kern-.1667em\lower.7ex\hbox{E}\kern-.125emX}}
\markboth{\journalname, VOL. XX, NO. XX, XXXX 2017}
{Author \MakeLowercase{\textit{et al.}}: Preparation of Papers for IEEE Control Systems Letters (August 2022)}

\title{Multi-Agent Temporal Logic Planning via Penalty Functions and Block-Coordinate Optimization}

\author{Eleftherios E. Vlahakis, \IEEEmembership{Member, IEEE}, Arash Bahari Kordabad, Lars Lindemann, \IEEEmembership{Member, IEEE},\\ Pantelis Sopasakis, Sadegh Soudjani, \IEEEmembership{Member, IEEE} and Dimos V. Dimarogonas, \IEEEmembership{Fellow, IEEE}
\thanks{This work was supported by the Swedish
Research Council, the Knut \& Alice Wallenberg Foundation, the Horizon Europe Grant SymAware, the ERC Consolidator Grant LEAFHOUND, and KK-stiftelsen under grant no. 202400088.}
\thanks{Eleftherios E. Vlahakis is with the Division of Computer Engineering and Computer Science, University West, 46132, Trollhättan, Sweden; \texttt{eleftherios.vlahakis@hv.se}}
\thanks{Arash Bahari Kordabad and Sadegh Soudjani are with the Max Planck Institute for Software Systems, Kaiserslautern, Germany; \texttt{\{arashbk, sadegh\}@mpi-sws.org}}
\thanks{Lars Lindemann is with the Automatic Control Laboratory, ETH Zürich, 8092, Zürich, Switzerland; \texttt{llindemann@ethz.ch}}
\thanks{Pantelis Sopasakis is with the School of Electronics, Electrical Engineering and Computer Science, Queen's University Belfast, Northern Ireland, BT9 5BN, UK; \texttt{p.sopasakis@qub.ac.uk}}
\thanks{Dimos V. Dimarogonas is with the Division of Decision and Control Systems, School of Electrical Engineering and Computer Science, KTH Royal Institute of Technology, 10044, Stockholm, Sweden; \texttt{dimos@kth.se}}
}

\maketitle
\thispagestyle{empty}

\begin{abstract}
Multi-agent planning under Signal Temporal Logic (STL) is often hindered by collaborative tasks  
that lead to computational challenges due to the inherent high dimensionality of the problem, 
preventing scalable synthesis with satisfaction guarantees. To address this, we formulate STL planning as an optimization program under multi-agent STL constraints and introduce a penalty-based unconstrained relaxation that can be efficiently solved via a Block-Coordinate Gradient Descent (BCGD) method, where each block corresponds to a single agent's decision variables, thereby mitigating complexity. By utilizing a quadratic penalty function defined via smooth STL semantics, we show that BCGD iterations converge to a stationary point of the penalized problem under standard regularity assumptions. To enforce feasibility,  
the BCGD solver is embedded within a 
two-layer optimization scheme: inner BCGD updates are performed for a fixed penalty parameter, which is then increased in an outer loop to progressively improve multi-agent STL robustness. The proposed framework enables scalable computations 
and is validated through various complex 
multi-robot planning scenarios.
\end{abstract}


\section{Introduction}
\label{sec:introduction}

\IEEEPARstart{M}{ulti-agent} systems (MAS) research deals with the task of coordinating  
collections of autonomous systems, e.g., in logistics, exploration, and smart infrastructure. These applications require agents to satisfy complex spatio-temporal and logical constraints governing both individual and interactive behaviors. Signal Temporal Logic (STL) \cite{MalerSTL2004} has emerged as a powerful framework suitable for these requirements, offering an expressive language to encode time-bounded properties over continuous-time signals.

Unlike automata-based LTL synthesis \cite{Belta2017formal}, STL's quantitative semantics \cite{Fainekos2009, Donze2010} enable the direct optimization of satisfaction margins over system trajectories. While exact solutions can be obtained via mixed-integer MPC formulations \cite{MurrayCDC2014, Kurtz2022}, their poor scalability has motivated the development of smooth robustness relaxations   
\cite{Pant2017, MehdipourACC2019, Gilpin2020, Tumova25} for efficient gradient-based optimization.

However, in MAS settings, collaborative tasks  
further amplify computational complexity, rendering scalable planning formidable. Important existing multi-agent approaches, including distributed MPC \cite{Charitidou2024, Zhou2022}, decentralized feedback control \cite{Lindemann2019auto}, and sequential planning \cite{VlahakisCDC24}, address these challenges but are often limited to restricted STL fragments, rely on the existence of feasible solutions, or employ heuristic coordination schemes.    
Developing scalable multi-agent planning methods that can handle complex collaborative tasks while maintaining computational tractability under a broad class of STL formulas    
thus remains an important open problem.

To address this challenge, this paper introduces a scalable, optimization-based framework for multi-agent STL planning under arbitrary collaborative tasks, leveraging smooth STL semantics~\cite{Gilpin2020} and the computational efficiency of the Block-Coordinate Gradient Descent (BCGD) method~\cite{Tseng2009}. Specifically, we demonstrate that the original planning problem, featuring an objective function that is separable across agents, yet subject to generally coupled multi-agent STL constraints, can be relaxed into an unconstrained problem via a quadratic penalty defined over smooth robustness metrics. This relaxation is solved efficiently via BCGD, where computations are performed at the block level, with each block corresponding to the decision variables of a single agent. 
Under standard regularity assumptions, we show that the BCGD iterations converge to a stationary point of the penalized problem, providing a computational architecture that remains tractable despite the complexity of the multi-agent specification. To enforce feasible solutions for the original planning problem, the BCGD solver is embedded in a two-layer optimization scheme, forming a penalty method (PM)~\cite[Chap.~17]{nocedal2006book}, where the inner loop optimizes for a fixed penalty parameter, which is then updated in the outer loop to progressively improve multi-agent STL robustness. 

This framework is the first to systematically integrate smooth STL semantics, block-coordinate optimization, and penalty functions for efficient multi-agent STL synthesis. We validate BCGD-PM across complex multi-robot scenarios, benchmarking it against an LBFGS-based implementation~\cite[Chap.~9]{nocedal2006book},  \cite[Sec.~7]{Tseng2009} within the same modular penalty framework. This comparison highlights BCGD's consistency in handling complex multi-agent STL planning. For readability, the main technical proofs are provided in the Appendix.

\section{Problem setup}\label{sec:Prob_setup}

\noindent\textbf{Notation: }
The sets of real numbers and nonnegative integers are $\R$ and $\N$, respectively. Let $N\in \N$ so that $\N_{[0,N]}=\{0,1,\ldots,N\}$. Let $x_{1},\ldots,x_{n}$ be vectors so that $x=(x_{1},\ldots,x_{n}) = [x_{1}^\intercal  \;\cdots\;x_{n}^\intercal  ]^\intercal$. Let $a,b\in\N$ with $a\leq b$. We denote by $\bm{x}(a{:}b)=(x(a),\ldots,x(b))$ an aggregate vector consisting of $x(t)$, $t\in \N_{[a,b]}$, representing a trajectory, or   
an aggregate trajectory when 
$x(t)=(x_1(t),\ldots,x_M(t))$ with $M\in \N$.  
The cardinality of a set $\cV$ is $|\cV|$.  
We call $x^\star$ a \emph{stationary point} of $f\,{:}\,\R^n {\to} \R$ if its gradient at $x^\star$ is 
$\nabla f(x^\star) {=} 0$.  
Given a sequence $\{x^k\}$,  
$\bar{x}$ is called an
\emph{accumulation point} if there exists a subsequence
$\{x^{k_j}\}$ such that $x^{k_j} {\to} \bar{x}$.

\subsection{Signal temporal logic}  
We consider STL formulas in positive normal form (PNF): 
\begin{equation}\label{eq:STL_syntax}
    \varphi 
    \coloneqq 
    \top 
    \mid
    \pi
    \mid
    {\lnot} \pi  
    \mid
    \phi_1 {\wedge} \phi_2
    \mid
    \phi_1 {\lor} \phi_2
    \mid
    \square_\cI \phi_1
    \mid
    \phi_1\, \cU_\cI\phi_2,
\end{equation}
where $\pi\coloneqq(\mu(x)\geq 0)$ is a predicate, with predicate function $\mu:\R^{n_x}\to \R$, $\phi_1$ and $\phi_2$ are STL formulas built recursively using the grammar in \eqref{eq:STL_syntax}, $\neg$, $\wedge$, and $\lor$ are the logical operators denoting \textit{negation}, \textit{conjunction}, and \textit{disjunction}, respectively, and $\square_{\cI}$ and $\cU_{\cI}$ are the \textit{always} and \textit{until} temporal operators, respectively, defined over the discrete interval $\cI{\subset} \N$. We omit the \textit{eventually} operator ($\lozenge_{\cI}$) from \eqref{eq:STL_syntax} since $\lozenge_{\cI} \phi_1 {=} \top \cU_{\cI}\phi_1$. 

The above definition has negation appearing only beside atomic predicates. This form of STL specifications in PNF is equivalent to the full class of STL specifications \cite{MalerSTL2004}, and any STL formula can be transformed into PNF using usual logical identities~\cite[Prop.~2]{Sadraddini2015}.

We denote by $\bm{x}(t) \models \phi$, $t\in\N$, the satisfaction 
of $\phi$, verified over $\bm{x}(t)=(x(t),x(t+1),\ldots)$. 
The validity of $\phi$ can be determined recursively using the Boolean semantics of STL; for details, we refer to \cite{MalerSTL2004} due to space limitations.

STL is endowed with quantitative semantics \cite{Donze2010}: The \emph{robustness function}  
$\rho^\phi: \bigtimes_{t=0}^{\cH^\phi} \R^n \to \R$ 
of a signal $\bm{x}(t)$,  
where $\cH^\phi$ is the \emph{horizon} of $\phi$~\cite{MalerSTL2004},   
indicates how robustly a signal $\bm{x}(t)$ satisfies a formula $\phi$, and 
is defined recursively as
\begin{align}\label{eq:robustness_function}
\rho^\pi(\bm{x}(t)) 
&{=} \mu(x(t)), \nonumber\\[0.em]
\rho^{\lnot \pi}(\bm{x}(t)) 
&{=} {-}\rho^{\pi}(\bm{x}(t)), \nonumber \\[0.em]
\rho^{\phi_1 \land \phi_2}(\bm{x}(t)) 
&{=} \min\!\big(\rho^{\phi_1}(\bm{x}(t)),\, \rho^{\phi_2}(\bm{x}(t))\big), \nonumber \\[0.em]
\rho^{\phi_1 \lor \phi_2}(\bm{x}(t)) 
&{=} \max\!\big(\rho^{\phi_1}(\bm{x}(t)),\, \rho^{\phi_2}(\bm{x}(t))\big),  \nonumber\\[0.em]
\rho^{\square_\cI \phi_1}(\bm{x}(t)) &{=} \min_{\tau\in t\oplus \cI} \rho^{\phi_1}(\bm{x}(\tau)),  \nonumber\\[0.em]
\rho^{\phi_1 \cU_{\cI} \phi_2}\mkern-2mu(\bm{x}(t)) 
&{=} \mkern-8mu\max_{\substack{\\[.1pt]\tau \in t \oplus \cI}}\mkern-7mu 
   \Big(\mkern-7mu \min\!\big( \rho^{\phi_2}\mkern-2mu(\bm{x}(\tau)), \mkern-17mu\min_{\substack{\\[.1pt]\tau' \in \N_{[t,\tau]}}}\mkern-17mu\rho^{\phi_1}\mkern-2mu(\bm{x}(\tau'))\big) \mkern-6mu  \Big)\mkern-3.mu,  
\end{align}
where $\oplus$ is the Minkowski sum.  
The satisfaction of  
$\phi$ by  
$\bm{x}(t)$ is indicated by  
its robustness function in the sense that $\rho^\phi(\bm{x}(t)){>}0\Rightarrow \bm{x}(t)\models \phi$, $\rho^\phi(\bm{x}(t)){<}0\Rightarrow \bm{x}(t)\not\models \phi$, while $\rho^\phi(\bm{x}(t)){=}0$ does not in general determine satisfiability.

\subsection{Multi-agent system}\label{sec:MAS}

\subsubsection{Dynamics} We consider a MAS with $M$ agents, with the $i^{\text{th}}$ agent following the dynamics
\begin{equation}\label{eq:individual_agent_dynamics}
    x_i(t+1)=f_i(x_i(t),u_i(t)),
\end{equation}
where $x_i(t)\in \R^{n_i}$ and $u_i(t)\in\R^{m_i}$ are the state and input vectors, respectively, $f_i:\R^{n_i}\times \R^{m_i}\to \R^{n_i}$ is continuously differentiable and locally Lipschitz in $(x_i,u_i)$, $t{\in}\N$, and the initial condition $x_i(0)$ is known,  with $i\in\cV{=}\{1,\ldots,M\}$, where $\cV$ is the set of all agents.  
To formally group agents participating in collaborative tasks, we adopt the notion of \emph{cliques} from graph theory, defined next \cite{Orlin1977}.
\begin{definition}\label{def:cliques}
    Consider an undirected graph $\cG = (\cV,\cE)$ potentially containing self-loops and multiple edges with node set $\cV$   
    and edge set $\cE$. Let $\nu\subseteq \cV$  
    and $\cE_\nu\subseteq \cE$ be the set of edges connecting the nodes in $\nu$. Then, $(\nu,\cE_\nu)$ is called a clique if $\forall\, \nu_i,\nu_j{\in}\nu$, $(\nu_i,\nu_j){\in}\cE_\nu$, i.e., it is a complete subgraph of $\cG$.
\end{definition}

Consider a graph $\cG = (\cV,\cE)$ with clique set $\cK$. With a slight abuse of notation, we denote a clique $(\nu,\cE_\nu) \in \cK$ simply by $\nu$, and write $|\nu|$ for the number of nodes in $\nu$.  
Let $\nu\in\cK$  contain the agents $i_1,\ldots,i_{|\nu|}$, i.e., $\nu=\{i_1,\ldots,i_{|\nu|}\}$.    
By collecting individual state and input vectors, as 
$x_\nu(t)=(x_{i_1}(t),\ldots,x_{i_{|\nu|}}(t))\in \R^{n_\nu}$ and 
$u_\nu(t)=(u_{i_1}(t),\ldots,u_{i_{|\nu|}}(t))\in\R^{m_\nu}$, respectively, we write the aggregate dynamics of $|\nu|$ agents as
\begin{equation}\label{eq:clique_dynamics}
    x_\nu(t+1) = f_\nu(x_\nu(t),u_\nu(t)),
\end{equation}
\sloppy
with $f_\nu(x_\nu(t),u_\nu(t))=\left(f_{j}(x_{j}(t),u_{j}(t))\right)_{j\in\nu}$.   
When $\nu=\cV$, the aggregate dynamics of the entire MAS are written as
\begin{equation}\label{eq:MAS}
    x(t+1)=f(x(t),u(t)),
\end{equation}
with $x(t)=(x_1(t),\ldots,x_M(t))$, $u(t)=(u_1(t),\ldots,u_M(t))$, and $f=(f_1, \ldots ,f_M)$.

\subsubsection{STL specification}  
The MAS is subject to  
\begin{equation}
    \phi = \bigwedge_{\nu \in \cK_\phi} \phi_\nu, \label{eq:global_phi}    
\end{equation}
which is a conjunctive STL formula, where each conjunct \(\phi_\nu\) is defined over $\bm{x}_\nu(t)$ and follows the syntax in \eqref{eq:STL_syntax}, with the aggregate trajectory $\bm{x}_\nu(t)$ collecting the individual trajectories of the agents in the clique $\nu$.   
The set \(\cK_\phi\) collects all these cliques induced by $\phi$, and may include individual agents $(|\nu|=1)$ or groups of agents $(1<|\nu|\leq |\cV|)$. Note that different cliques may overlap in their agent sets, indicating that some agents participate in multiple collaborative tasks.

Let \(\pi {\coloneqq} (\mu(y) {\geq} 0)\) be a predicate in \(\phi\), where \(\mu: \R^{n_y} {\to} \R\). We assume that all predicate functions $\mu$ appearing in $\phi$ are continuously differentiable. The vector \(y {\in} \R^{n_y}\) may represent an individual state \(x_i {\in} \R^{n_i}\), for \(i {\in} \cV\), an aggregate state \(x_\nu {\in} \R^{n_\nu}\) for a clique \(\nu {\in} \cK_\phi\), or an aggregate state $x_\kappa{\in}\R^{n_\kappa}$ collecting the states of a subset of agents in a clique $\nu$.   
For example, in formula 
$\phi_\nu {=} \pi_\nu^{\kappa_1} \cU_{\cI} \pi_\nu^{\kappa_2}$, where $\nu=\{1,2,3\}$, the predicate $\pi_\nu^{\kappa_1} {\coloneqq} (x_1 {\geq} 0)$ involves one agent $\kappa_1 {=} \{1\}$,   
while $\pi_\nu^{\kappa_2} {\coloneqq} (x_2 {-} x_3 {\geq} 0)$ involves two agents $\kappa_2 {=} \{2,3\}$.

\subsection{Problem statement}
The multi-agent STL planning synthesis problem is formulated as an optimization problem over the multi-agent control sequence $\bm{u} {=} (u(0),\ldots,u(N{-}1))$, where $u(t){=}(u_1(t), \ldots, u_M(t))$, and $N{=}\cH^\phi$. Given the initial condition $x(0){=}x_0$ and the dynamics $x(t{+}1) {=} f(x(t), u(t))$, the multi-agent trajectory $\bm{x}{=}(x(0),\ldots,x(N))$, where $x(t){=}(x_1(t), \ldots, x_M(t))$, is explicitly determined by $\bm{u}$. We thus denote the multi-agent STL robustness $\rho^\phi(\bm{x})$ as $\rho^\phi(\bm{u})$, and the total cost as $\cL(\bm{u}) {=} \sum_{i\in\cV} \cL_i(\bm{u}_i)$, where $\cL_i(\bm{u}_i) = \sum_{t=0}^{N-1} \ell_i(x_i(t), u_i(t), t) + V_{f,i}(x_i(N))$, with $\bm{u}_i {=} (u_i(0), \dots, u_i(N-1))$ denoting the $i^{\text{th}}$ agent's inputs.   
Function $\ell_i:\R^{n_i}\times \R^{m_i}\times \N\to \R$ is the running cost for the $i^{\text{th}}$ agent, penalizing, e.g., state energy and control effort quantities that cannot be directly encoded through the specification $\phi$. The terminal cost $V_{f,i}:\R^{n_i}\to \R$
penalizes deviations from a prescribed terminal condition, which can be chosen, e.g., to enforce horizon-end requirements in $\phi$ or to induce cyclic trajectory planning by penalizing the distance $x_i(N){-}x_i(0)$ for all $i\in\cV$.    
The multi-agent STL planning problem is formulated as 
\begin{subequations}
\label{eq:multi_agent_problem}
    \begin{align}
    &\operatorname*{Minimize}_{\bm{u}} \,\, \cL(\bm{u}) = \sum_{i\in\cV} \cL_i(\bm{u}_i) \label{eq:cost_multi_agent_problem} \\
    &\text{subject\;to} \,\, \rho^\phi(\bm{u}) = \min_{\nu\in\cK_\phi} \rho^{\phi_\nu}(\bm{u}_\nu) > 0. \label{eq:joint_STL_constraints}
\end{align}
\end{subequations}
Solving \eqref{eq:multi_agent_problem} 
is computationally challenging due to the non-smoothness and coupling in the joint STL constraint in \eqref{eq:joint_STL_constraints}, which implies that $\bm{x}_\nu{\models} \phi_\nu$, $\forall \nu{\in}\cK_\phi$, i.e., $\bm{x}{\models} \phi$. 
We address this challenge by employing a smooth STL approximation and   
a penalty-based block-coordinate gradient descent method, which enables agent-level computations, mitigating the   
complexity of the centralized problem. This approach relies on the following assumption. 
\begin{assumption}\label{ass:Li} 
    The function $\cL_i:\R^{n_{\bm{u}_i}}\to \R$, $i\in\cV$, is proper, convex, continuous,
    and level-bounded.  
    Furthermore, the problem in 
    \eqref{eq:multi_agent_problem} is feasible.
\end{assumption}
In the remainder of the paper, Assumption~\ref{ass:Li} holds. We note that the first part
holds for  
linear dynamics, 
or, more generally, when the 
running cost $\ell_i$ depends only on the 
input variables and 
when 
the 
terminal cost $V_{f,i}$ is omitted. To streamline the presentation of the proposed optimization framework, we omit   
explicit input constraints from \eqref{eq:multi_agent_problem}.

\section{Multi-agent STL optimization}\label{sec:STLoptimization}

\subsection{Smooth STL approximations}

We recall that the STL constraint in  
\eqref{eq:joint_STL_constraints}  
is non-smooth, involving $\min$ and $\max$ operations over predicate functions. We underapproximate $\min$ and $\max$ operators \cite{Gilpin2020} as
\begin{subequations}\label{eq:min_max_approx}
\begin{align}
& \min \left(\mu_1,\ldots,\mu_q\right) 
\overset{^{\geq}}{\approx} 
-\frac{1}{\Gamma} \log\left(\sum_{j=1}^q e^{-\Gamma \mu_j}\right),
\label{eq:softmin}\\
& \max \left(\mu_1,\ldots,\mu_q\right) 
\overset{^{\geq}}{\approx} 
\frac{\sum_{j=1}^q \mu_j e^{\Gamma \mu_j}}{\sum_{j=1}^q e^{\Gamma \mu_j}},
\label{eq:softmax}
\end{align}
\end{subequations}
where the approximation becomes tighter for larger values of $\Gamma {>} 0$. For a fixed $\Gamma$, we denote by $\varrho^\phi_\Gamma(\bm{u})$ the smooth approximation of the robustness function $\rho^\phi(\bm{u})$, which satisfies $\varrho^\phi_\Gamma(\bm{u}) {\leq} \rho^\phi(\bm{u})$ for all $\Gamma {>} 0$, since $\phi$ is in PNF \cite{Kazemi2020}. This implies that the STL constraint in \eqref{eq:joint_STL_constraints}   
can be replaced by $\varrho^\phi_\Gamma(\bm{u}) {>} 0$, introducing additional conservatism in satisfying $\bm{x} \models \phi $  
controlled by $\Gamma$.  
In fact, due to the multi-agent structure of $\phi$ in \eqref{eq:global_phi}, the strict requirement $\varrho^\phi_\Gamma(\bm{u}) {>} 0$ can be relaxed to $\varrho^\phi_\Gamma(\bm{u}) {\geq} 0$, as formally stated next. 

\begin{proposition}\label{prop:varrho}
Consider the formula $\phi$ in~\eqref{eq:global_phi}, where $\cK_\phi$ contains at least two cliques, i.e., $|\cK_\phi| \geq 2$. Then, for all multi-agent sequences $\bm{u}$ and  $\Gamma > 0$, the smooth robustness satisfies $\varrho_\Gamma^\phi(\bm{u}) {<} \rho^\phi(\bm{u})$. Consequently, 
\(
    \varrho_\Gamma^\phi(\bm{u}) {\geq} 0 \Rightarrow \rho^\phi(\bm{u}) {>} 0.
\)
\end{proposition}

\subsection{Unconstrained STL optimization}

Due to Proposition \ref{prop:varrho}, a stricter version of the problem in \eqref{eq:multi_agent_problem} can now be written   
using smooth STL semantics as  
\begin{align}\label{eq:compact_problem_tight}
    \operatorname*{Minimize}_{\bm{u} \in \cM} \, \cL(\bm{u}), \; \text{where}\; \cM \coloneqq \{\bm{u} \mid \varrho_\Gamma^\phi(\bm{u}) \geq 0\}.
\end{align}
By Assumption \ref{ass:Li}, this problem has a minimizer \cite{nocedal2006book}.  
We address this constrained program using a suitable penalty function $R(\bm{u})$ leading to an unconstrained problem:  
\begin{align}\label{eq:unconstrained_problem_tight}
    \operatorname*{Minimize}_{\bm{u}}\, F_\lambda(\bm{u})
    \coloneqq\cL(\bm{u})+\lambda R(\bm{u}),
\end{align}
for some $\lambda{>}0$. To specify a suitable penalty $\lambda R(\bm{u})$ that relaxes the original constrained formulation into the unconstrained problem in \eqref{eq:unconstrained_problem_tight}   
we introduce the quadratic penalty 
\begin{align}\label{eq:penalty_function}
    R(\bm{u}) \coloneqq \max\{0, -\varrho_\Gamma^\phi(\bm{u})\}^2,
\end{align}
which satisfies $\cM = \{\bm{u} \mid R(\bm{u}) = 0\}$, and $R(\bm{u}) {>} 0$ for all $\bm{u} {\notin} \cM$.   
This is a  
differentiable penalty function with gradient
\begin{equation}
    \nabla R(\bm{u}) = -2\max\{0, -\varrho_\Gamma^\phi(\bm{u})\}\nabla\varrho_\Gamma^\phi(\bm{u}),
\end{equation}
which exists everywhere and is continuous. 
We shall first discuss how the unconstrained problem 
in \eqref{eq:unconstrained_problem_tight} can be solved 
for a fixed value of $\lambda > 0$ using a block coordinate 
gradient descent method. Subsequently, we will show how a 
penalty method can be used to solve the  
constrained problem in \eqref{eq:compact_problem_tight}.

\subsection{Block coordinate gradient descent}
Problem \eqref{eq:unconstrained_problem_tight} possesses a 
certain structure that can be exploited to solve it 
efficiently. First, note that the cost function $\cL(\bm{u})$  
is separable into agent-level objective functions $\cL_i(\bm{u}_i)$, which by Assumption~\ref{ass:Li} are proper, convex, and continuous in $\bm{u}_i$ for all $i \in \cV$. The penalty term $\lambda R(\bm{u})$, which penalizes violations of the smooth robustness condition $\varrho_\Gamma^\phi(\bm{u}) \ge 0$, is differentiable.
These properties allow us to use the block coordinate gradient descent (BCGD) method \cite{Tseng2009}. 
In each BCGD iteration, (i) $\lambda R(\bm{u})$ is approximated by a strictly convex quadratic function, enabling the application of block coordinate descent to generate a descent direction, and (ii) a sufficient descent direction is computed for an agent-level block of $\bm{u}$, as detailed in Algorithm~\ref{alg:BCGD}.

\smallskip
\noindent\textbf{Quadratic approximation:}
We model the variation $R(\bm{u}^k{+}\bm{d}) -R(\bm{u}^k)$   
at $\bm{u}^k$  
by $Q^{H}(\bm{u}^k, \bm{d}) \coloneqq \nabla R(\bm{u}^k)^\intercal \bm{d} + \tfrac{1}{2} \bm{d}^\intercal H^k \bm{d}$, 
where $\bm{d}$ is the 
update direction to $\bm{u}^k$ derived from \eqref{eq:update_direction}, and $H^k {\succ} 0$ approximates the Hessian $\nabla^2 R(\bm{u}^k)$.

\smallskip
\noindent\textbf{Block selection:}
In iteration $k$ we choose a nonempty subset of agent-level blocks $J^k \subseteq \cV$,
so that only the elements $(\bm{u}_i)_{i\in J^k}$ are updated.
Over iterations, all blocks should be updated at least once (generalized Gauss--Seidel), or one may select the most ``active" blocks using a Gauss--Southwell type rule for efficiency---see \cite{Tseng2001,Tseng2009} for details.

\smallskip
\noindent\textbf{Block update direction:}
We compute the descent direction
\begin{equation}
    \bm{d}^k \mkern-2mu {=}\mkern-2mu 
    \argmin_{\bm{d}} \mkern-4mu\Big\{ \lambda Q^{H}\mkern-3mu(\bm{u}^k\mkern-3mu, \bm{d})  {+} \cL(\bm{u}^k {+} \bm{d}) {\Big |} \bm{d}_j {=} 0, \forall j {\notin} J \mkern-2mu\Big\}.\label{eq:update_direction}
\end{equation}
Due to the separability of $\cL(\bm{u})$, this step decomposes into independent subproblems for each selected agent block when $H^k$ is chosen block-diagonal,
i.e., it is equivalent to solving
\begin{equation}
    \operatorname*{Min.}_{(\bm{d}_j)_{j\in J^k}} 
   \mkern-3mu \sum_{j\in J^k} \mkern-3mu
    \frac{\lambda}{2} \bm{d}_j^\intercal H_j^k\bm{d}_j 
    {+} \lambda \nabla R(\bm{u}^k)_j^\intercal \bm{d}_j 
    {+} \mathcal{L}_j(\bm{u}_j^k{+}\bm{d}_j).
\end{equation}
Thus, for $j\in J^k$, we have the optimization problems
\begin{equation}
    \operatorname*{Min.}_{\bm{d}_j}
    \,
    \frac{\lambda}{2} \bm{d}_j^\intercal H_j^k\bm{d}_j
    + \lambda \nabla R(\bm{u}^k)_j^\intercal \bm{d}_j
    + \mathcal{L}_j(\bm{u}_j^k+\bm{d}_j)
,
    \label{eq:update_direction:subproblem}
\end{equation}
which are convex and can be solved efficiently, where $H_j^k {\succ} 0$ are the diagonal blocks of $H^k$, enabling a parallel computation across all agents in $J^k$, when $|J^k|{\ge}1$. Matrix $H^k$ approximates the Hessian $\nabla^2 R(\bm{u}^k)$, which is costly to compute. In fact, any $H_j^k \succ 0$ may be used, with sufficient decrease along $\bm{d}_j^k$ ensured via the inexact line search below.

\smallskip
\noindent\textbf{Armijo-type rule for step-size selection:} 
Let $\sigma, \gamma \in (0, 1)$, $\Delta F_\lambda^k = F_\lambda(\bm{u}^k {+} \alpha^k \bm{d}^k) - F_\lambda(\bm{u}^k)$, and $\Delta \cL^k = \cL(\bm{u}^k {+}   
\bm{d}^k) - \cL(\bm{u}^k)$. 
We select the largest $\alpha^k {\in} \{1/ 2^j\}_{j\in\N}$ to ensure sufficient decrease of $F_\lambda(\bm{u})$ along $\bm{d}^k$, i.e.,
\begin{equation}\label{eq:armijo}
   \Delta F_\lambda^k 
    \leq 
    \sigma\alpha^k\mkern-0mu 
    \left(  \lambda\nabla R(\bm{u}^k)^\intercal \bm{d}^k 
    {+}
    \gamma (\bm{d}^k)^\intercal H^k \bm{d}^k {+} \Delta \cL^k \right).
\end{equation}

\noindent\textbf{Initialization rule:} 
Let $\cT_i = \{\nu \in \cK_\phi \mid i \in \nu \}$ be the set of cliques containing $i$. Assuming  
$\{i\}{\in} \cT_i$ for all $i{\in}\cV$, implying that for all $i{\in}\cV$ there is an individual task $\phi_i$, one can obtain initial guesses for $\bm{u}_i^0$ by minimizing \(
    \cL_i(\bm{u}_i)
    {+} \lambda\, \max\{0,\, -\varrho_\Gamma^{\phi_i}(\bm{u}_i)\}^2\) over $\bm{u}_i$, for $i{\in}\cV$, 
    using, e.g.,  
    the  
    toolbox in \cite{open2020}.  
    Otherwise, one may start with $\bm{u}^0{=}0$.

\smallskip
\noindent\textbf{Termination:}
Following \cite{Tseng2009}, Alg.~\ref{alg:BCGD}
stops when $\|\mkern-2mu H^k \bm{d}^k\mkern-2mu\|_\infty \mkern-2mu{\leq} \epsilon$.

\begin{algorithm}[t]
\caption{Block Coordinate Gradient Descent (BCGD)}
\begin{algorithmic}[1]
\State \textbf{Input:} 
$\bm{u}^0$ (initial iterate),  
$\lambda$ (penalty), 
$K$ (maximum iterations),  
$\epsilon>0$ (tolerance)
\For{$k = 0,1,\dots,K-1$}
   \State Select agent blocks $J^k \subseteq \cV$ and an $H^k\succ 0$ 
    \State Solve \eqref{eq:update_direction} with $H{=}H^k$ and $J=J^k$ to obtain $\bm{d}^k$ 
    \State \textbf{If} $\|H^k \bm{d}^k\|_\infty \leq \epsilon$ \textbf{break}
    \State Choose step size $\alpha^k$ and set $\bm{u}^{k+1} = \bm{u}^k + \alpha^k \bm{d}^k$
\EndFor
\State \textbf{return} $\bm{u}^{\star, k}$
\end{algorithmic}\label{alg:BCGD}
\end{algorithm}

\begin{proposition}
\label{prop:bcgd}
For a fixed penalty parameter $\lambda{>}0$, let $\{\bm{u}^k\}$ be the sequence
generated by Alg.~\ref{alg:BCGD} applied to~\eqref{eq:unconstrained_problem_tight}. 
Then every accumulation point of $\{\bm{u}^k\}$ is a stationary point of \eqref{eq:unconstrained_problem_tight}.
\end{proposition}

The proof of Proposition \ref{prop:bcgd} follows directly from 
\cite[Theorem 1]{Tseng2009} tailored to the multi-agent STL setting considered here.   
For fixed $\lambda$, the objective $F_\lambda$   
consists of a proper, convex, continuous block-separable cost and a smooth quadratic penalty function. Consequently, the assumptions of~\cite{Tseng2009} are satisfied, and the proposed BCGD scheme converges to a stationary point of $F_\lambda$.  
The most computationally demanding part of Alg.~\ref{alg:BCGD}
is computing the block elements $\nabla R(\bm{u}^k)_j$, followed by the computation 
of the block update direction $\bm{d}^k$.    
Next, we embed the BCGD algorithm in a penalty framework.

\subsection{Penalty method for the constrained problem}

Here, we propose a penalty method \cite[Chap. 17]{nocedal2006book} for the constrained problem \eqref{eq:compact_problem_tight}.  
Specifically, we solve a sequence of unconstrained problems of the form \eqref{eq:unconstrained_problem_tight}, each corresponding to a fixed penalty parameter $\lambda >0$, and iteratively increase $\lambda$ to progressively enforce feasibility. Following \cite[Alg. 1]{open2020}, 
the quadratic penalty method is given in Alg. \ref{alg:penalty_method}.   
Due to the nonconvex nature of the problem,
no theoretical guarantees of convergence to a global 
optimum can be provided. However, as we will demonstrate, in practice
Alg.~\ref{alg:penalty_method} can yield an approximate solution $\bar{\bm{u}}^\star$ that is $\epsilon_{\rm infeas}$-infeasible, satisfying $\varrho_\Gamma^\phi(\bar{\bm{u}}^\star) {>} {-}\sqrt{\epsilon_{\rm infeas}}$.
Due to the true--smooth STL semantics gap (Prop.~\ref{prop:varrho}), such near-feasible solutions may satisfy the original (non-smooth) STL constraints \eqref{eq:joint_STL_constraints}.
This behavior is formalized as a conditional guarantee in Theorem~\ref{thm:penalty}.

\begin{algorithm}[t]
\caption{Penalty method (PM)}
\begin{algorithmic}[1]
\State \textbf{Input:} $\bm{u}^0$ (initial iterate), 
$\lambda^0$ (initial penalty), 
$K_{\rm PM}$ (maximum iterations), 
$\eta_\lambda > 1$ (penalty update factor), 
$\epsilon_{\rm infeas} > 0$ (max. infeasibility), 
$\epsilon^0$ (initial tolerance),
$\eta_\epsilon \in (0, 1]$ (tolerance update parameter)
\State $\bar{\bm{u}}^0 = \bm{u}^0$
\For{$k = 0,1,\dots,K_{\rm PM}-1$}
    \State \textbf{if} $R(\bar{\bm{u}}^{k}) < \epsilon_{\rm infeas}$ \textbf{break} 

    \State Use Alg. \ref{alg:BCGD} to solve \eqref{eq:unconstrained_problem_tight} with $\lambda = \lambda^k$, initial guess $\bar{\bm{u}}^k$, and tolerance $\epsilon^k$; obtain $\epsilon^k$-approx. solution $\bm{u}^{\star, k}$
    \State $\lambda^{k+1} = \eta_\lambda \lambda^k$, $\bar{\bm{u}}^{k+1}=\bm{u}^{\star, k}$, $\epsilon^{k+1} = \eta_\epsilon \epsilon^k$
    
\EndFor
\State \textbf{return } $\bar{\bm{u}}^{\star,k}$ 
\end{algorithmic}\label{alg:penalty_method}
\end{algorithm}

\begin{theorem}\label{thm:penalty}
Suppose that $|\cK_\phi| {\geq} 2$ and $\bm{u}^0$ is such that the sequence $\{\bar{\bm{u}}^{k}\}$ 
of Alg.~\ref{alg:penalty_method} has a limit point $\bar{\bm{u}}^\star$ 
with $R(\bar{\bm{u}}^\star) {=} 0$.
For a given problem of the form \eqref{eq:compact_problem_tight}, there is $\epsilon_{\rm infeas}^\star$
such that Alg. \ref{alg:penalty_method} with $\epsilon_{\rm infeas} {\leq} \epsilon_{\rm infeas}^\star$ 
returns a feasible solution for 
\eqref{eq:multi_agent_problem}.
\end{theorem}

Although the existence of a limit point, $\bar{\bm{u}}^\star$, in Theorem~\ref{thm:penalty} is not guaranteed to be met under nonlinear 
dynamics and general multi-agent STL constraints, 
Section~\ref{sec:example} demonstrates the effectiveness of the proposed 
approach in multi-robot STL planning. 
Moreover, the ``inner'' optimization problems \eqref{eq:unconstrained_problem_tight} are solved
efficiently with Alg. \ref{alg:BCGD}, while Alg. \ref{alg:penalty_method} requires only a small number of ``outer'' iterations to converge. 

\section{Numerical validation via ten-robot example}\label{sec:example}

\begin{figure*}[]
    \centering
    \input{figures/RURAMCA_unic_fig_final}
    \caption{\textbf{RURAMCA} motion planning for ten robots with discrete-time unicycle dynamics:
\(z_i(t+1)=z_i(t)+v_i(t)\cos\theta_i(t)\),
\(y_i(t+1)=y_i(t)+v_i(t)\sin\theta_i(t)\),
and \(\theta_i(t+1)=\theta_i(t)+\omega_i(t)\),
where \((z_i(t),y_i(t))\in\R^2\) denotes the Cartesian position of the $i^{\text{th}}$ robot,
\(\theta_i(t)\in\R\) its heading, and
\(v_i(t)\in\R\), \(\omega_i(t)\in\R\) the linear and angular control inputs, respectively, with a sampling period of $1$ s. The trajectories are generated from initial conditions (colored squares) using input sequences obtained via Alg. \ref{alg:penalty_method} (using Alg.~\ref{alg:BCGD} as inner solver). Simulation setup: $\Gamma{=}2$ for inner smooth operators, $\Gamma{=}1$ for the outer softmin in \eqref{eq:joint_STL_constraints}, $\sigma{=}0.5$ and $\gamma{=}0.995$ (Armijo rule), $\bm{u}^0{=}0$, $\lambda^0{=}1$, $\eta_\lambda{=}5$, $\epsilon_{\rm infeas}{=}5.0{\times}10^{-4}$, $\epsilon{=}10^{-6}$, $H^k{=}10^3 I$, $\ell_i{=}v_i(t)^2{+}\omega_i(t)^2$ for all $t{\in}\cI$, and $V_{f,i}{=}0$. Block direction update: 
$\bm{d}_i^k {=} {-}(10^3\lambda^k {+} 2)^{-1}
\left(2\bm{u}_i^k {+} \lambda^k\nabla R(\bm{u}^k)_i\right)$.
$x_i(0)$ (\(\square\)), $x_i(100)$ (\pentagon), $i{\in}\N_{[1,10]}$ (filled colored).}
    \label{fig:ten_robot_determ_traj}
\end{figure*}

We evaluate the proposed framework on the ten-robot planning problem in \cite{VlahakisCDC24, VlahakisCDC25}, considering the multi-agent specification \(\phi{=}\bigwedge_{\nu\in\cK_\phi}{\phi_\nu}\) with horizon \(\cH^\phi{=}100\) over \(\cI{=}\N_{[0,100]}\), where \(\cK_\phi\) contains 21 cliques.   
Each task $\phi_\nu$ is either individual or collaborative. For $\nu=\{i\}$, the $i^{\text{th}}$ robot must avoid obstacles ($\square_{\cI}\neg \cO_l$), visit $C_i$ within $\cI_c=\N_{[10,50]}$ ($\lozenge_{\cI_c}C_i$), and reach $D_i$ within $\cI_d=\N_{[70,100]}$ ($\lozenge_{\cI_d}D_i$). The workspace with regions and obstacles is shown in Fig.~\ref{fig:ten_robot_determ_traj}. For $|\nu|>1$, robots in clique $\nu$ must meet ($\lozenge_{\cI_m}M_\nu$) within $\cI_m{=}\N_{[0,70]}$. The formulas $C_i$, $D_i$, $i{\in}\cV$, and $\cO_l$, $l{\in}\N_{[1,3]}$, are 
conjunctions of predicates, each defined by a linear predicate function, while 
$M_\nu{\coloneqq} (\min_{\kappa\subseteq\nu}\mu_\kappa(x_\kappa){\geq} 0)$, where $x_\kappa{=}(z_q,y_q)_{q\in\kappa}$, with $\kappa{=}(i,j){\subseteq} \nu$ being any subset of two robots in $\nu$, and $\mu_\kappa(x_\kappa) = 0.25 - \bigl\| [I \; {-I}]\, x_\kappa \bigr\|$. This scenario is termed R2AM.

We consider two additional planning scenarios building on the R2AM baseline:  
R2AMCA, which augments R2AM with global inter-agent collision-avoidance 
specified as $\square_{\cI} \bigwedge_{i\neq j}(\|x_i(t){-}x_j(t)\|{\geq} 0.01)$, and RURAMCA, which connects the two reach specifications (R2) in R2AMCA (or R2AM) with the \emph{until} operator specified as 
$(\lozenge_{\cI_c}C_i) \,\cU_{[0,50]} (\lozenge_{\cI_c} D_i)$. 
Moreover, robots must fulfill their specific collaborative meeting tasks, $\lozenge_{\cI_m} M_\nu$, while ensuring collision-free motion as in the R2AMCA plan.

Motivated by \cite{Tseng2009}, we compare BCGD with LBFGS~\cite[Chap.~9]{nocedal2006book} implemented in a block-coordinate manner (\texttt{jaxopt.LBFGS}), highlighting the modularity of the proposed penalty framework. Both solvers use identical randomized shuffling, updating each block once every 10 iterations. Table~\ref{tab:runtime_comparison} reports runtime, robustness, and feasibility statistics over 100 runs with permuted initial conditions from Fig.~\ref{fig:ten_robot_determ_traj}.

Our numerical results show that BCGD consistently achieves feasible solutions across all scenarios under both linear and unicycle dynamics. 
Although LBFGS is faster on average (when considering only those problems 
that it can solve), BCGD is both significantly 
more robust (it solves all problems) and has a lower 95\%-quantile 
solve time---cf. Table~\ref{tab:runtime_comparison}.
These observations are consistent with the benchmarking in \cite{Tseng2009}, albeit outside an STL context.

\begin{table*}[]
\centering
\setlength{\tabcolsep}{3pt}
\renewcommand{\arraystretch}{0.95}
\caption{Runtime (s), robustness ($\rho \times 10^3$), and feasibility statistics over 100 randomized permutations of the agents' initial conditions in Fig.~\ref{fig:ten_robot_determ_traj}. Entries report mean $\pm$ std (robustness in parentheses), with feasibility rates in brackets. L and U denote linear and unicycle dynamics. For each scenario, the solver achieving the lower 95\%-quantile runtime is highlighted in bold.}
\label{tab:runtime_comparison}
\resizebox{\textwidth}{!}{%
\begin{tabular}{llll}
\toprule
\textbf{Solver} & \textbf{R2AM} & \textbf{R2AMCA} & \textbf{RURAMCA} \\ 
\midrule
BCGD (L) 
& \textbf{2.43$\pm$0.43} (13.88$\pm$11.72) [100]
& \textbf{2.99$\pm$1.29} (10.18$\pm$10.36) [100]
& \textbf{14.07$\pm$4.02} (6.55$\pm$6.49)  [100]\\

LBFGS (L) 
& 1.62$\pm$4.96 (23.13$\pm$86.51) [92]
& 4.19$\pm$21.69 (0.66$\pm$17.55) [68]
& 10.23$\pm$16.31 (1.62$\pm$8.91) [75]\\
\midrule
BCGD (U) 
& \textbf{182.86$\pm$138.07} (9.36$\pm$14.48) [100]
& \textbf{223.44$\pm$130.38} (5.58$\pm$6.23) [100]
& \textbf{337.31$\pm$83.81} (3.23$\pm$3.44) [100]\\
LBFGS (U) 
& 83.77$\pm$295.29 (4.46$\pm$18.00) [88]
& 97.72$\pm$311.45 (3.46$\pm$8.24) [83]
& 180.05$\pm$303.65 (2.42$\pm$16.88) [80] \\
\bottomrule
\end{tabular}%
}
\end{table*}

\begin{figure}
    \centering
    \begin{tikzpicture}
\begin{groupplot}[
    group style={
        group size=1 by 2,
        vertical sep=0.1cm,
    },
    width=0.44\textwidth,
    height=0.15\textwidth,
    ymode=log,
    ylabel={Runtime (s)},
    boxplot/draw direction=y,
    xtick={1,2,3},
    xticklabels={
    \shortstack{$N{=}100$\\$10$ agents},
    \shortstack{$N{=}500$(U), $10^3$(L)\\$10$ agents},
    \shortstack{$N{=}100$\\$20$ agents}},
    grid=major,
    major grid style={dotted,gray!50},
]

\nextgroupplot[
    title={Unicycle (U)},
    title style={yshift=-.75cm, xshift=-1.35cm},xtick=\empty,
    xticklabels=\empty,ylabel = \empty,ymode=log
]

\addplot+[
    boxplot={draw position=1, box extend=0.25},
] table[row sep=\\, y index=0] {
110.61\\291.36\\62.91\\243.86\\255.79\\431.31\\160.73\\163.98\\456.85\\74.05\\114.71\\103.5\\117.08\\91.29\\328.51\\57.36\\173.32\\309.97\\169.22\\35.73\\
244.81\\117.51\\101.66\\279.11\\62.48\\161.46\\60.56\\73.8\\445.02\\116.02\\53.36\\78.71\\70.17\\79.48\\263.91\\203.92\\499.56\\530.17\\130.82\\62.97\\
601.59\\765.31\\45.44\\141.69\\63.58\\217.45\\97.93\\154.61\\514.42\\494.89\\291.87\\104.92\\184.42\\130.05\\238.43\\287.88\\128.09\\408.39\\146.88\\139.06\\
132.54\\83.93\\67.66\\137.43\\113.64\\118.88\\158.42\\256.14\\79.41\\108.29\\58.67\\189.21\\117.26\\215.26\\62.8\\172.42\\489.1\\115.34\\78.74\\69.16\\
75.13\\126.32\\111.34\\194.63\\69.4\\96.28\\74.37\\201.22\\107.34\\139.44\\62.12\\116.52\\141.42\\152.2\\259.86\\220.01\\127.76\\308.58\\157.13\\175.68\\
};

\addplot+[
    boxplot={draw position=2, box extend=0.25},
] table[row sep=\\, y index=0] {
2193.62\\2614.76\\1909.77\\664.03\\565.75\\551.91\\1138.54\\409.07\\472.21\\829.32\\1592.15\\779.48\\1675.51\\320.11\\1281.1\\859.64\\1535.54\\1121.13\\1078.89\\1177.22\\
208.39\\1803.86\\1060.12\\1797.47\\516.62\\352.1\\2543.26\\257.33\\1175.17\\1420.11\\263.12\\1927.15\\1786.45\\777.59\\1139.95\\937.8\\780.12\\2045.27\\1040.39\\1027.53\\
1339.27\\253.81\\372.88\\1438.89\\695.21\\1686.73\\428.81\\847.66\\1153.51\\2477.66\\1368.99\\1330.24\\2728.2\\1228.64\\1374.33\\898.12\\494.43\\789.92\\197.2\\1024.85\\
1598.53\\845.94\\617.87\\997.69\\729.34\\1224.94\\215.21\\633.71\\404.79\\1243.13\\1288.86\\1876.06\\1798.21\\1323.84\\2580.4\\819.86\\689.01\\1377.22\\2819.15\\636.98\\
1656.53\\2030.37\\498.44\\578.9\\414.39\\1007.42\\1174.47\\368.29\\1204.48\\1797.94\\1417.84\\672.96\\1554.21\\631.17\\2316.83\\1861.39\\962.86\\412.04\\412.49\\2625.54\\
};

\addplot+[
    boxplot={draw position=3, box extend=0.25},
] table[row sep=\\, y index=0] {
1645.29\\1490.61\\1127.32\\743.17\\1567.27\\1770.67\\981.53\\1013.76\\1103.89\\861.13\\1187.85\\784.04\\1217.82\\705.15\\1460.5\\818.47\\1260.35\\1096.53\\1104.25\\999.07\\
844.34\\954\\964.65\\886.49\\849.98\\855.42\\1151.99\\1432.11\\1316.84\\1206.78\\1396.78\\1060.78\\933.42\\1656.82\\904.25\\1596.38\\1420.43\\811.68\\1719.28\\904.19\\
1120.95\\951.48\\934.31\\884.08\\1268.97\\868.7\\1015.6\\687.46\\1256.55\\1121.46\\694.62\\831.62\\1491.57\\887.68\\658.31\\1186.98\\853.7\\1285.96\\951.95\\1462.96\\
814.93\\1209.66\\1228.6\\998.85\\749.63\\686.2\\970.66\\1034.13\\876.77\\1219.94\\1213.51\\1107.2\\989.74\\619.21\\1086.85\\876.34\\674.73\\1027.46\\997.32\\1178.59\\
845.59\\962.32\\741.29\\1343.26\\922.56\\882.2\\1081.8\\1327.91\\1193.01\\970.73\\935.94\\963.35\\1417.43\\1335.7\\1142.55\\957.18\\1351.59\\1658.3\\726.34\\1416.91\\
};

\nextgroupplot[
    title={Linear (L)},
    title style={yshift=-.75cm, xshift=-1.35cm},
    ylabel={Runtime (s)},
    ylabel style={xshift=0.6cm, yshift=.cm},ymode=linear
]
\addplot+[
    boxplot={draw position=1, box extend=0.25},
] table[row sep=\\, y index=0] {
2.56\\1.89\\1.93\\3.3\\2.51\\2.44\\2.42\\2.53\\2.56\\2.48\\1.96\\1.9\\2.23\\2.39\\2.17\\2.05\\2.59\\3.06\\2.95\\2.21\\
2.53\\1.62\\2.31\\2.96\\2.61\\3.36\\2.14\\1.87\\2.81\\2.43\\1.68\\2.2\\2.18\\1.99\\2.46\\3.06\\2.79\\2.66\\2.83\\2.53\\
2.55\\2.96\\1.95\\2.14\\2.01\\2.66\\1.87\\2.41\\2.9\\3.14\\2.91\\1.89\\2.68\\3.41\\3.58\\1.79\\2.29\\2.34\\2.16\\1.75\\
2.17\\2.2\\1.83\\2.46\\2.29\\2.93\\2.6\\3.03\\2.19\\2.85\\2.17\\1.85\\2.78\\2.66\\2.17\\2.68\\2.88\\2.26\\2\\2.14\\
3.05\\2.37\\1.95\\3.06\\2.08\\1.91\\2.21\\2.31\\2.28\\2.12\\1.93\\2.1\\2.5\\2.6\\2.83\\2.85\\1.94\\3.01\\2.57\\2.94\\
};

\addplot+[
    boxplot={draw position=2, box extend=0.25},
] table[row sep=\\, y index=0] {
14.24\\11.05\\9.28\\26.23\\9.39\\17.45\\10.78\\10.67\\17.87\\14.32\\10.16\\4.01\\14.8\\16.08\\13.03\\11.64\\15.71\\19.51\\16.32\\6.01\\
17.2\\11.88\\14.56\\19.18\\15.87\\27.42\\10.48\\12.06\\24.44\\16.1\\5.97\\14.22\\8.32\\9.63\\16.64\\22.09\\24.99\\23.17\\16.41\\11.34\\
17.3\\22.18\\7.88\\13.86\\10.07\\27.27\\8.71\\16.22\\25.34\\20.33\\24.23\\13.86\\20.59\\27.14\\34.37\\7.45\\13.29\\12.62\\15.23\\10.83\\
18.95\\11.76\\8.36\\12.27\\12.87\\17.81\\12.88\\20.95\\10.08\\16.99\\5.8\\10.34\\20.36\\15.12\\13.11\\17.68\\25.46\\13.64\\10.11\\11.88\\
17.25\\11.93\\12.6\\22.75\\10.34\\10.03\\10.92\\17.4\\13.67\\13.08\\12.13\\15.83\\24.98\\14.14\\18.86\\17.44\\12.63\\16.35\\12.99\\24.56\\
};

\addplot+[
    boxplot={draw position=3, box extend=0.25},
] table[row sep=\\, y index=0] {
7.01\\9.89\\5.82\\8.31\\8.6\\11.31\\11.79\\6.4\\12.22\\11.66\\10.36\\4.91\\10.77\\11\\7.98\\4.82\\7.42\\6.66\\11.74\\6.6\\
9.14\\8.16\\6.07\\8.88\\7.03\\16.23\\7.59\\10.58\\12.77\\9.82\\5.96\\9.4\\8.65\\8.25\\10\\13.29\\12.96\\5.89\\10.56\\9.6\\
7.51\\12.38\\9.15\\11.11\\11.58\\11.48\\6.54\\8.36\\10.56\\12.24\\9.75\\7.82\\8\\11.94\\10.37\\4.51\\9.98\\8.58\\9.38\\7.07\\
15.02\\7.2\\14.26\\8.05\\8.35\\10.14\\9.64\\11.74\\15.27\\5.61\\10.53\\13.53\\8.34\\12.02\\9.06\\11.56\\7.72\\8.89\\9.05\\7.87\\
7.87\\10.04\\7.08\\11.45\\8.47\\7.93\\9.81\\10.25\\10.76\\13.33\\6.24\\10.37\\13.43\\10.3\\6.95\\9.39\\7.54\\9.85\\10.91\\6.37\\
};

\end{groupplot}
\end{tikzpicture}
    \caption{Runtime boxplots for scalability of BCGD-PM over 100 randomized initial-condition permutations.}
    \label{fig:scalability}
\end{figure}

To assess scalability, we evaluate BCGD under increasing horizons and problem sizes (see Fig. \ref{fig:scalability} for runtime statistics). Runtime scales gracefully with the horizon (e.g., $10\times$ increase yields $\sim\!6\times$ runtime under linear dynamics) while maintaining feasibility and similar robustness. Doubling agents and cliques leads to subquadratic-to-moderate runtime growth with no loss in feasibility and minor robustness variation. These results indicate favorable scalability of the block-coordinate structure in the reported experiments, particularly under linear dynamics, while remaining reliable for nonlinear models. Additionally, BCGD-PM significantly outperforms our earlier methods in \cite{VlahakisCDC24,VlahakisCDC25} in runtime for R2AM, whereas in R2AMCA and RURAMCA these methods fail to find feasible solutions within a $10^4$\,s timeout. The proposed algorithm is implemented in Python using \href{https://github.com/jax-ml/jax}{JAX}. Source code and animations are available at \href{https://github.com/lefterisvl83/MAS-STL-planning}{github.com/lefterisvl83/MAS-STL-planning}.

\section{Conclusions}
\label{sec:concl}

We presented a scalable framework for multi-agent planning under collaborative STL tasks. We employed a penalty method to tackle the underlying optimization problem under smooth STL semantics, using unconstrained optimization with quadratic penalty functions. We showed that the structure of the resulting unconstrained subproblems admits an efficient block-coordinate gradient descent solution, enabling agent-level computations with convergence guarantees while mitigating computational complexity. 
The proposed framework motivates future research into reactive STL planning for time-varying MAS structures and 
tasks.

\section*{Appendix: proofs}

\noindent\textbf{Proposition~\ref{prop:varrho}:}
Let $a_\nu = \rho^{\phi_\nu}(\bm{u}_\nu)$ for all $\nu \in \cK_\phi$, and let
\(
a_{\min} = \min_{\nu \in \cK_\phi} a_\nu = \rho^\phi(\bm{u})
\)
denote the true robustness. By definition, the smooth robustness  
is 
\(
\varrho_\Gamma^\phi(\bm{u})
= -\frac{1}{\Gamma}
\log \sum_{\nu \in \cK_\phi} e^{-\Gamma a_\nu}.
\)
Writing each $a_\nu$ as $a_\nu = a_{\min} + (a_\nu - a_{\min})$ and factoring
out $e^{-\Gamma a_{\min}}$ from the summation yields
\(
\sum_{\nu \in \cK_\phi} e^{-\Gamma a_\nu}
= e^{-\Gamma a_{\min}}
\sum_{\nu \in \cK_\phi} e^{-\Gamma (a_\nu - a_{\min})}.
\)
Substituting this expression back into the definition of
$\varrho_\Gamma^\phi(\bm{u})$ gives
\(
\varrho_\Gamma^\phi(\bm{u})
= -\frac{1}{\Gamma}
\log\!\left(
e^{-\Gamma a_{\min}}
\sum_{\nu \in \cK_\phi} e^{-\Gamma (a_\nu - a_{\min})}
\right) = a_{\min}
- \frac{1}{\Gamma}
\log \sum_{\nu \in \cK_\phi} e^{-\Gamma (a_\nu - a_{\min})}.
\)
Defining
\(
\delta_\Gamma(\bm{u})
:= \frac{1}{\Gamma}
\log \sum_{\nu \in \cK_\phi} e^{-\Gamma (a_\nu - a_{\min})},
\)
we obtain 
\(
\varrho_\Gamma^\phi(\bm{u}) = a_{\min} - \delta_\Gamma(\bm{u}).
\) 
Since $a_\nu \ge a_{\min}$ for all $\nu \in \cK_\phi$, it follows that
$e^{-\Gamma (a_\nu - a_{\min})} \le 1$. Because at least one term in $\sum_{\nu \in \cK_\phi} e^{-\Gamma (a_\nu - a_{\min})}$  is 1, each $a_\nu$ is finite by  continuity of the system dynamics and the predicate functions, and  $|\cK_\phi| \ge 2$ by assumption, the summation
inside the logarithm satisfies
\(
1 < \sum_{\nu \in \cK_\phi} e^{-\Gamma (a_\nu - a_{\min})}
\le |\cK_\phi|,
\)
which implies
\(
0 < \delta_\Gamma(\bm{u})
\le \frac{\log |\cK_\phi|}{\Gamma}\) for any \(\Gamma > 0,
\)
hence, 
\(
\varrho_\Gamma^\phi(\bm{u}) = a_{\min}-\delta_\Gamma(\bm{u}) < \rho^\phi(\bm{u}).
\)
Therefore,
\(
\varrho_\Gamma^\phi(\bm{u}) {\ge} 0
{\Rightarrow}
\rho^\phi(\bm{u}) {>} 0
\).

\noindent\textbf{Theorem~\ref{thm:penalty}:} 
Let us define the continuous function $\delta_\Gamma(\bm{u}) := \rho^\phi(\bm{u}) - \varrho^\phi_\Gamma(\bm{u}) > 0$
(from Prop.~\ref{prop:varrho}),
so for every compact set $\mathbb{U}\subset \R^{n_{\bm{u}}}$, 
$\alpha_\mathbb{U} \coloneqq \min_{\bm{u}\in \mathbb{U}}\delta_\Gamma(\bm{u}) > 0$.  
Note that for $\mathbb{U} \supseteq \mathbb{U}'$
it holds that $\alpha_{\mathbb{U}} \leq \alpha_{\mathbb{U}'}$. 
Let $\mathcal{N}(\bar{\bm{u}}^\star)$ be a compact neighborhood of 
$\bar{\bm{u}}^\star$.
We have
$\delta_\Gamma(\bm{u}) \geq \alpha_{\mathcal{N}(\bar{\bm{u}}^\star)}$
for all $\bm{u} \in \mathcal{N}(\bar{\bm{u}}^\star)$.
By the assumption on the existence of a limit point
$\bar{\bm{u}}^\star$ such that $R(\bar{\bm{u}}^\star)=0$,  
there is a subsequence
$\{\bar{\bm{u}}^{\star, \iota_s}\}_s$ which converges to $\bar{\bm{u}}^{\star}$,
so there is $s \in \N$ so that 
$\bar{\bm{u}}^{\star, \iota_s} \in \mathcal{N}(\bar{\bm{u}}^\star)$
and, by continuity of $R$, $R(\bar{\bm{u}}^{\star, \iota_s}) < \alpha_{\mathcal{N}(\bar{\bm{u}}^\star)}^2$. Therefore,
$\varrho^\phi_\Gamma(\bar{\bm{u}}^{\star, \iota_s}) > -\alpha_{\mathcal{N}(\bar{\bm{u}}^\star)}$, hence 
$\rho^\phi(\bar{\bm{u}}^{\star, \iota_s}) > 0$.

\bibliographystyle{IEEEtran}
\bibliography{biblio}

\end{document}